\documentclass[12pt,english,a4paper]{article}
\usepackage[english,activeacute]{babel}
\usepackage[T1]{fontenc}
\usepackage{graphicx}
\usepackage[section]{placeins}
\usepackage{setspace}
\usepackage{amsmath}
\usepackage{amsfonts}
\usepackage{amssymb}
\usepackage{array}
\usepackage{color}
\usepackage{colortbl}
\usepackage[small,bf]{caption}
\usepackage{multirow}
\usepackage[affil-it]{authblk}
\usepackage[utf8]{inputenc}
\usepackage{authblk}
\usepackage[all]{xy}

\begin{document}

\title{ Molecular Dissociation in Presence of a Catalyst II: The bond
breaking role of the transition from virtual to localized states.}
\author[1,2]{ A. Ruderman}
\author[3]{A. D. Dente}
\author[1,2]{E. Santos}
\author[1]{H. M. Pastawski}

\affil[1]{Instituto de F\'{i}sica Enrique Gaviola (CONICET), Facultad de
Matem\'{a}tica, Astronom\'{i}a y F\'{i}sica, Universidad Nacional de
C\'{o}rdoba, 5000, C\'{o}rdoba, Argentina}

\affil[2]{Institute of Theoretical Chemistry, Ulm University, D-89069, Ulm,
Germany}

\affil[3]{INVAP S.E., 8403, San Carlos de
Bariloche, Argentina}

\maketitle
\begin{abstract}
We address a molecular dissociation mechanism that is known to occur when a H%
$_{2}$ molecule approaches a catalyst with its \textit{molecular axis
parallel to the surface}. It is found that molecular dissociation is a form
of quantum dynamical phase transition associated to an analytic
discontinuity of quite unusual nature: the molecule is destabilized by the
transition from non-physical virtual states into actual localized states.
Current description complements our recent results for a molecule approaching
the catalyst with its \textit{molecular axis perpendicular to the surface} \cite{grado1}.
 Also, such a description can be seen as a further successful
implementation of a non-Hermitian Hamiltonian in a well defined model.
\end{abstract}

\section{Introduction}

How do molecules form? This has been recognized as one of the ten unsolved
mysteries of Chemistry, enumerated in 2013 for the Year of Chemistry
Celebration \cite{SciAm}. Indeed, a new entity emerges when two identical
atoms meet. The reciprocal is also true: as a dimer approaches a catalyst's
surface, it may break down. But when and how does this break down precisely
happen? What distinguishes these two different quantum objects, i.e. the
molecule and the two independent atoms? It is natural to think that as some
control parameter move, e.g. an inter-atomic distance, a sort of
discontinuity or phase transition should happen. While a quantum calculation
can be set up to simulate such reaction, the calculations of an increasingly
realistic system quickly begin to overwhelm even the most powerful computer.
While DFT calculations hint a change in chemical bonds as the molecule-catalyst interaction
increases when the molecule approaches to the
surface \cite{BONDBREACK-Santos-2011-Diatomic-molecules}, this is confronted
with the fact that in a finite system no actual discontinuities can happen.
The key for the molecule formation/dissociation mystery can be found in P.
W. Anderson's inspiring paper \textquotedblleft More is
Different\textquotedblright\ \cite{anderson1972more}. There, Anderson
recalled that the inversion oscillations in ammonia-like molecules suffer a
sort of transition into a non-oscillating mode as the masses are increased.
Much as in a classical oscillator transition to an over-damped regime, the
crucial ingredient is the infinite nature of the environment which induces
dissipation while preventing the occurrence of Poincare's recurrences and
enable a dynamical phase transition. These concepts were formalized in the
context of the Rabi oscillations in a quantum system: a spin dimer immersed
in an environment of spins. Since this is solved in the thermodynamic limit
of infinitely many spins which provide the crucial continuum spectrum. \cite%
{w1954mathematical,Alvarez-LevsteinJCP2006}. In this case, the finite Rabi
frequency undergoes a non-analytic transition into a non-oscillatory mode as
the interaction with the environment increases \cite{sachdev2007quantum,
RedutionismBerry,leggett1987dynamics}. This mathematical discontinuity was
termed Quantum Dynamical Phase Transition (QDPT) \cite{Pastawski2007278}.

While the application of these ideas to molecular dissociation/formation is
not completely straightforward, in a previous paper we succeeded in
describing H$_{2}$ molecule formation/dissociation in the presence of a
catalyst as a QDPT \cite{grado1}. This description was achieved using a
variant of the model introduced by D. M. Newns for hydrogen adsorption in a
metallic surface \cite{NewnsHydrogen-and-dband}. However, that analysis was
restricted to the case when \textit{the molecular axis is perpendicular to
the catalyst surface}. In \cite{grado1} the environment provides the
infinitely many catalyst orbitals whose influence had to be treated beyond
linear response. Indeed, the interaction among the crystal states and the
dimer orbitals dramatically perturb each other and has to be obtained
through a self-consistent Dyson equation. In particular, the substrate
induces imaginary corrections to the molecular energies, accounting for
their finite lifetime. These complex energies, as those obtained from the
Fermi Golden Rule, represent resonances and are accounted by a non-Hermitian
Hamiltonian \cite{rotter2015review}. Our main result was that two \textit{%
resonances are formed inside} the $d$ band and that they present analytical
discontinuities as a function of the molecule-substrate interaction
(distance) \cite{Berry,Rotter1,Rotter2}. Thus, the \textit{molecular
dissociation/formation} was identified as the non-analytic \textit{%
collapse/splitting of these resonances}.

In this paper, we address another reaction mechanism that is known to occur
when a H$_{2}$ molecule approaches a catalyst with its \textit{molecular
axis parallel to the surface}. It is found that molecular dissociation is
also a phase transition associated to an analytic discontinuity, but of
different and unusual nature: the molecule is destabilized by the transition
from non-physical virtual states into actual localized states. For the rest
of the article we will be dealing with the same model and tools introduced
in our previous work \cite{grado1} which, in this case, provide
substantially new perspective into the molecular dissociation/formation
problem.

\section{The model}

\begin{figure}[tbh!]
\centering{\ \includegraphics[width=5cm]{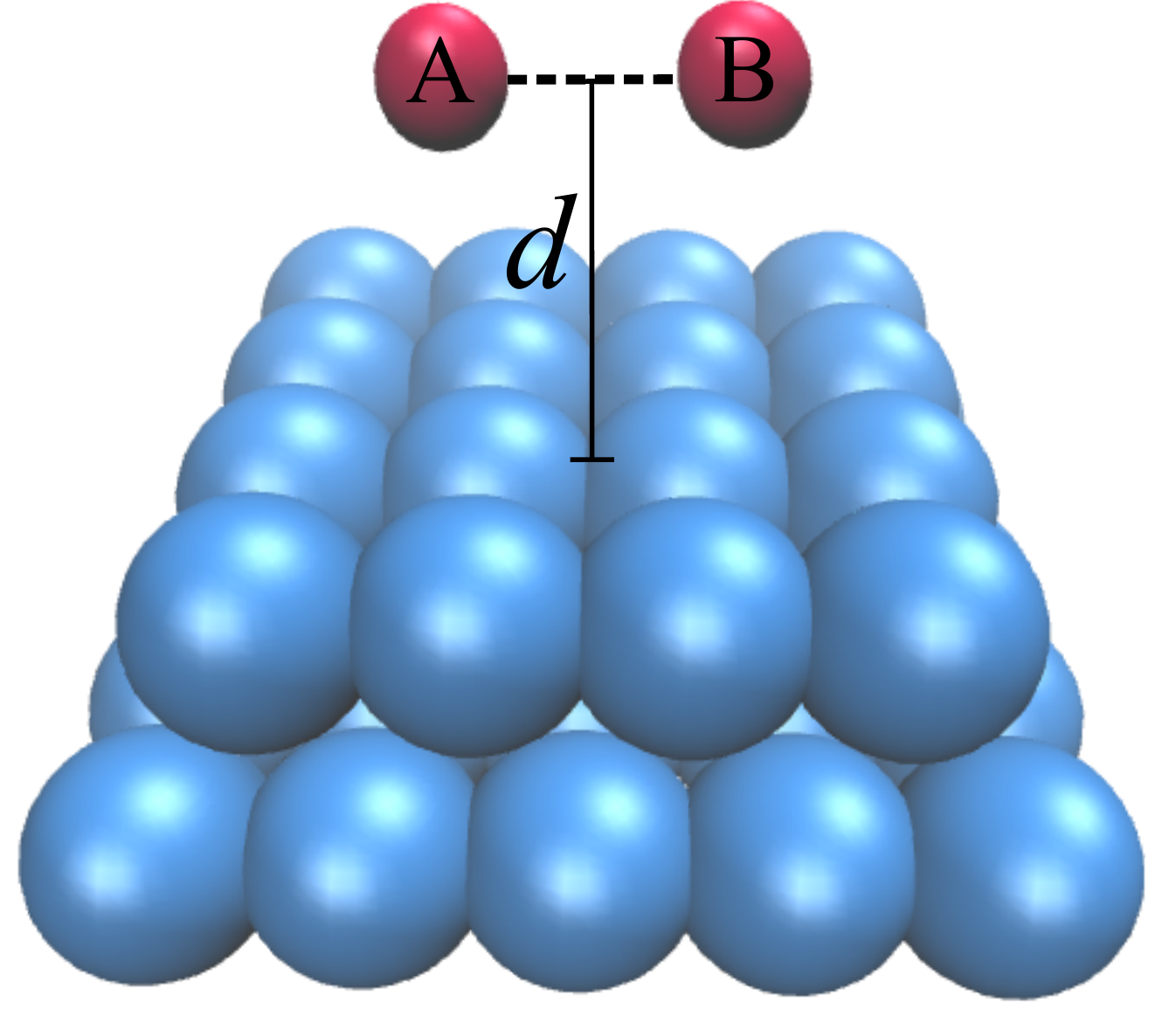} }
\caption{Homonuclear molecule interacting with a metallic surface. The
principal axis of the molecule is parallel to the surface and the distance
of each atom to the substrate are the same.}
\label{esquema}
\end{figure}

Given a homonuclear molecule AB and a metal electrode with a
half filled $d$ band, two independent geometries arise to describe the
interaction. The particular configuration of a molecule approaching with its
axis perpendicular to the metal surface, was previously investigated in
reference \cite{grado1}. A fully different problem arises when the axis
along the molecule lies parallel to the surface. In this configuration the
distances between a given atom belonging to the metal surface and both atoms
forming the molecule remain equal, i.e. $d_{\text{A}}=d_{\text{B}}=d$ ( see
Fig. \ref{esquema}). Therefore, both atoms interact identically with the
metal, resulting in a completely different Hamiltonian respect to the
perpendicular case, and hence yielding a dissimilar kind of transition.

\begin{figure}[tbh!]
\centering{\ \includegraphics[width=8cm]{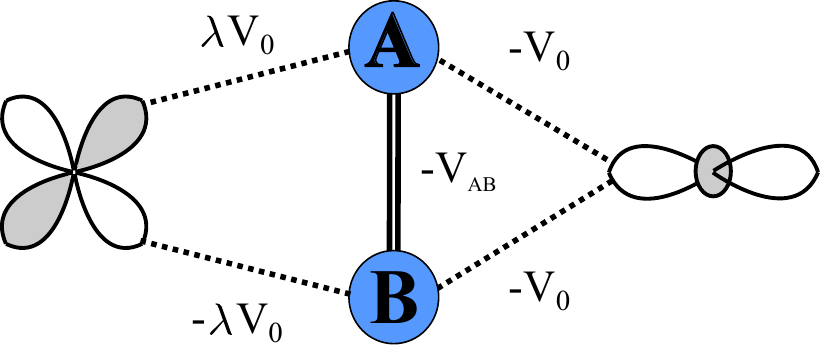} }
\caption{Different signs for the interaction between the molecule and the
metallic atomic orbital, due to the lobe phase shift for the atomic orbital
functions $d_{z^2}$ and $d_{xz}$. The $\protect\lambda$ factor accounts for
the different strength interaction between the molecule and the orbitals $%
d_{z^2}$ and $d_{xz}$.}
\label{sitios}
\end{figure}

To set up the model Hamiltonian for the interaction between the molecule and
the metal, we write the molecule's Hamiltonian as: 
\begin{equation*}
\hat{H}_{\mathrm{mol}}=E_{A}\left\vert A\right\rangle \left\langle
A\right\vert +E_{B}\left\vert B\right\rangle \left\langle B\right\vert
-V_{AB}\left( \left\vert A\right\rangle \left\langle B\right\vert
+\left\vert B\right\rangle \left\langle A\right\vert \right) .
\end{equation*}%
The atomic energies $E_{A}$ and $E_{B}$ are identical and their degeneracy
is broken by the mixing element $-V_{AB}$ that leads to the bonding and
antibonding states, i.e. the Highest Occupied Molecular Orbital (HOMO) and
Lowest Unoccupied Molecular Orbital (LUMO), respectively. In this
orientation, the molecule only can have substantial overlap with the metal $%
d_{z^{2}}$ and $d_{xz}$ orbitals of the underlying metallic atom. Therefore, 
$z$ is considered to be perpendicular to the surface and $x$ is chosen
parallel to the molecular axis. Both orbitals interact with the target
molecule in different ways \cite{Hoffman}, as depicted in Fig. \ref{sitios}.
On one side, the overlap of the $d_{z^{2}}$ with the atomic orbitals A and B
have the same the sign and magnitude, resulting in a Hamiltonian coupling
element $-V_{0}$. On the other side, the molecule also interacts with the $%
d_{xz}$ orbital of the metal. In this case, while having equal strengths a
different sign appears for each atomic orbital. Taking these considerations
into account, there are two concurrent mechanisms for molecule-metal
interaction : 
\begin{equation*}
\hat{V}_{\mathrm{int}}=V_{0}(\left\vert A\right\rangle \left\langle
d_{z^{2}}\right\vert +\left\vert B\right\rangle \left\langle
d_{z^{2}}\right\vert )+\lambda V_{0}(-\left\vert A\right\rangle \left\langle
d_{xz}\right\vert +\left\vert B\right\rangle \left\langle d_{xz}\right\vert
),
\end{equation*}%
where $\left\vert d_{z^{2}}\right\rangle $ and $\left\vert
d_{xz}\right\rangle $ are the metallic orbitals that interact with the
molecule. Furthermore, we have included a $\lambda $ factor to account for
the difference among the interaction strengths with the two $d$ orbitals.

\begin{figure}[tbh!]
\centering{\ \includegraphics[width=8cm]{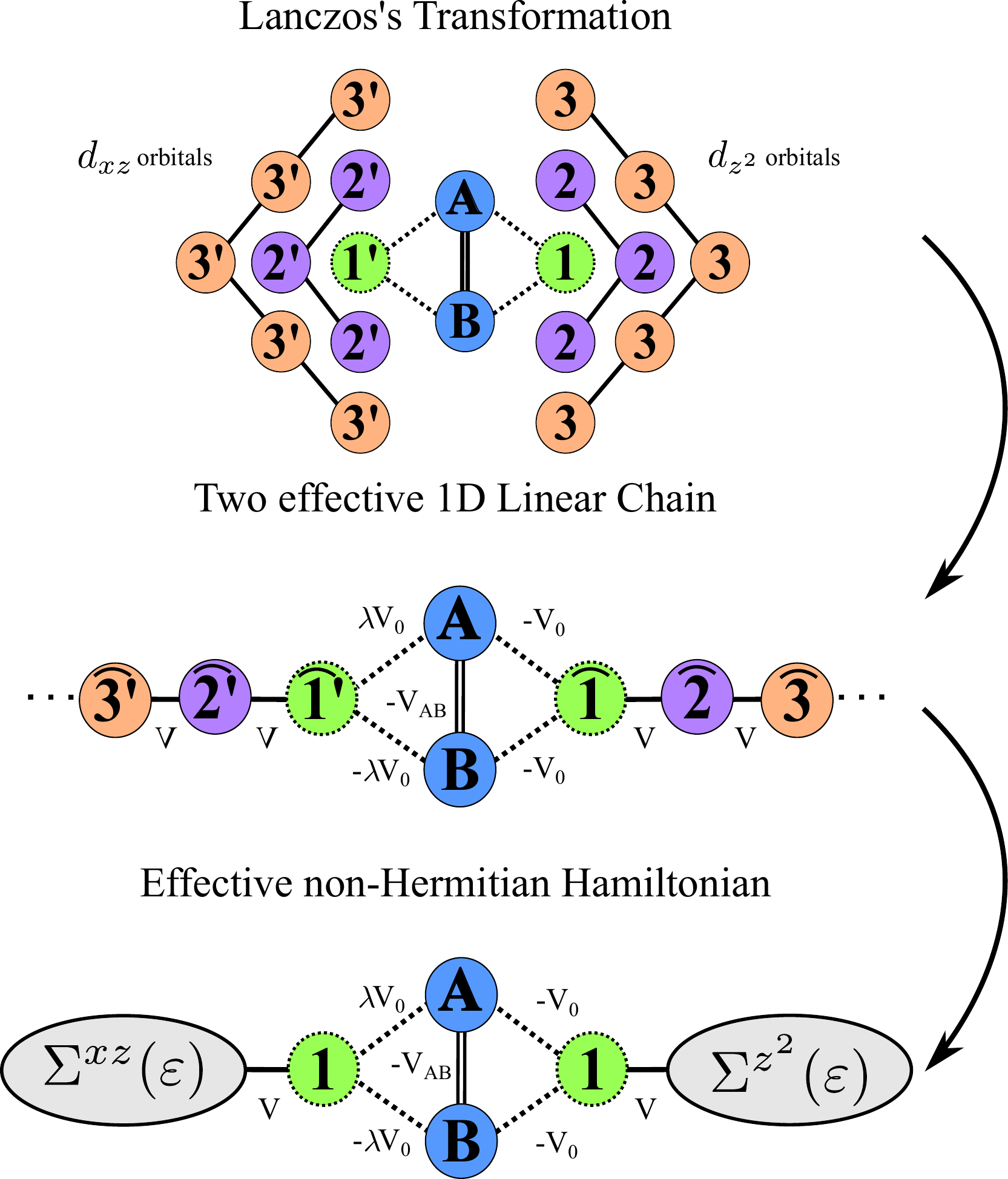} }
\caption{Effective non-Hermitian Hamiltonian due Lanczos's transformation
from a molecule A-B (in blue), interacting with a 2D metal substrate
composed of two distinct collections of $d$ orbitals. The transformation
implies combining each layer of orbitals at the same distance of the
interacting atom. The decimation process results in a four dimensional
Hamiltonian with the metal represented as two effective self-energies.}
\label{lanczos}
\end{figure}

In Fig. \ref{sitios} we represent explicitly two, assumed independent, sets
of metallic $d$ orbitals associated with each symmetry of the surface
orbitals (i.e. $\left\vert d_{z^{2}}\right\rangle $ and $\left\vert
d_{xz}\right\rangle $). Therefore, the relevant part of the metal
Hamiltonian can be represented using a narrow band model. This approximation
was first proposed by Newns \cite{NewnsHydrogen-and-dband}, who stated that
the projection of the $d$ band Local Density of States (LDoS) over the
specific orbital (either $d_{z^{2}}$ or $d_{xz}$) could be schematized as a
semielliptic energy band that strongly interacts with the molecule \cite%
{xin2014effects}. This picture is validated by appealing to a Lanczos's
transformation \cite{grado1,lanczos1950iteration,haydock1972electronic} to
obtain this simple electronic structure for the $d$ band. The basic
procedure is visualised in Fig. \ref{lanczos} for a two dimensional metal
represented as two distinct collections of orthonormal $d$ orbitals. By
choosing one of the interacting metallic orbitals as a reference, the
intermetallic interactions provide (through the Lanczos's procedure) for
combination of atomic $d$ orbitals consistent with the initial symmetry.
Typically, these are progressively included according to their distance to
the initial orbital. These \textquotedblleft collective\textquotedblright\
substrate orbitals are naturally arranged in the Hilbert space in order to
evidence the tridiagonal nature of the Hamiltonian in the new basis. By
means of this procedure, the general three dimensional geometry of a
catalyst is reduced to a effective linear chain. The same reasoning applies
for both symmetries. Then, we can write the metal $d_{z^{2}}$ Hamiltonian as:

\begin{equation}
\hat{H}_{\mathrm{met}}^{z^{2}}=\sum_{n=1}^{\infty }E_{n}^{z^{2}}\left\vert
n\right\rangle \left\langle n\right\vert -\sum_{n=1}^{\infty
}V_{n,n+1}^{z^{2}}\left( \left\vert n\right\rangle \left\langle
n+1\right\vert +\left\vert n+1\right\rangle \left\langle n\right\vert
\right) ,  \label{emet}
\end{equation}%
where $\left\vert n\right\rangle $ and $E_{n}^{z^{2}}$ are the $n$-th
collective metal orbital obtained by the Lanczos's transformation and the
energy corresponding to that orbital, respectively. For the sake of
simplicity, all the hopping elements $V_{n,n+1}^{z^{2}}$ are considered to
be equal to $V$. This is consistent with the fast convergence of the hopping
elements, first addressed by Haydock et al. \cite{haydock1972electronic}. A
similar Hamiltonian $\hat{H}_{\mathrm{met}}^{xz}$ is obtained for the $xy$
symmetry. Thus, 
\begin{equation*}
\hat{H}_{\mathrm{met}}^{{}}=\hat{H}_{\mathrm{met}}^{z^{2}}+\hat{H}_{\mathrm{%
met}}^{xz}.
\end{equation*}

In order to obtain an optimal configuration for our discussion on the
dissociation process \cite{Hush}, we make the $d$ band to be centered around
the Fermi energy $E$ by making $E_{A}=E_{B}=E_{n}=E$. Then, the bonding and
antibondig molecular states, i.e. HOMO and LUMO, fall outside the band as $%
2|V_{AB}|>4|V|$ \cite{Santos2011314}. This choice is consistent with the
standard knowledge of the Markus-Hush theory for optimal conditions of
electron transfer and molecular dissociation. In this work, we used $%
V_{AB}/V=2.5$ which is typical for H$_{2}.$

The main features of the system, i.e. energy spectrum and relevant
eigenvalues properties, could be obtained using a decimation procedure \cite%
{Levstein-Damato,pastawski-medina}. This formulation deploys an infinite
order perturbation theory for the interaction $\hat{V}_{\mathrm{int}}$ to
dress the molecular Hamiltonian $\hat{H}_{\mathrm{mol.}}$ into an effective
molecular Hamiltonian that accounts for the presence of the catalyst, and
yields a complex correction, $\Sigma $, to the molecular bonding and
antibonding energies. This is sketched in the bottom panel of Fig. \ref%
{lanczos}. This precisely defined procedure resorts to the Green's function
matrix associated with the total Hamiltonian $\hat{H}=\hat{H}_{\mathrm{mol}}+%
\hat{H}_{\mathrm{met}}+\hat{V}_{\mathrm{int}}$, 
\begin{equation}
\mathbb{G(}\varepsilon )=(\varepsilon \mathbb{I}-\mathbb{H})^{-1}.
\label{green}
\end{equation}%
We are going to profit from the fact that the poles of the Green's function
are the eigenvalues of the system. At this point, a brief introduction to
the decimation technique is convenient for the sake of clarity. Let us first
consider the molecular Hamiltonian without the presence of the metal: 
\begin{equation}
\mathbb{H}_{\mathrm{mol}}=\left[ 
\begin{array}{cc}
E_{A} & -V_{AB} \\ 
-V_{AB} & E_{B}%
\end{array}%
\right] .  \label{e1}
\end{equation}%
Then, the Green's function matrix adopts the form: 
\begin{equation}
\mathbb{G}_{\mathrm{mol}}=\dfrac{1}{(\varepsilon -E_{A})(\varepsilon
-E_{B})-|V_{AB}|^{2}}\left[ 
\begin{array}{cc}
\varepsilon -E_{B} & V_{AB} \\ 
V_{AB} & \varepsilon -E_{A}%
\end{array}%
\right] .  \label{molg}
\end{equation}%
The Green's function for atom A, the first diagonal element of $\mathbb{G}_{%
\mathrm{mol}}$, can be written as 
\begin{equation*}
G_{\mathrm{mol}}^{AA}=(\varepsilon -E_{A}-\Sigma _{A})^{-1}
\end{equation*}%
Therefore, the energy of atom A is modified by the presence of the atom B
through the self-energy 
\begin{equation*}
\Sigma _{A}=|V_{AB}|^{2}/(\varepsilon -E_{B}).
\end{equation*}%
This decimation procedure can be extended to the full semi-infinite chain
that describes the components of the $d$ band that couple with the HOMO and
LUMO according to their symmetry. The procedure consists on
\textquotedblleft dressing\textquotedblright\ the successive
\textquotedblleft Lanczos's orbitals\textquotedblright\ with the
corresponding self-energies to account for the interaction with the
neighbour atom at the right. In a finite system of $N+2$ orbitals, $\Sigma
_{A}$ is written in terms of $N+1$ levels of a continued fraction until one
reaches the last level. To simplify the study of the spectral density, the
energies of the system can be renormalized by introducing an imaginary small
quantity $-\mathrm{i}\eta $, thus $E\rightarrow E-\mathrm{i}\eta $. This
energy correction can be seen as a weak environmental interaction, a role
that could be assigned to the $sp$ band states \cite{CattenaBustosPRB2010}.
Thus, in the thermodynamic limit of a semi-infinite chain ($N\rightarrow
\infty $), the self-energy correction due to the metal becomes:

\begin{equation}
\Sigma (\varepsilon )=\dfrac{|V|^{2}}{\varepsilon -\left( E-\mathrm{i}\eta
\right) -\Sigma (\varepsilon )}=\Delta (\varepsilon )-\mathrm{i}\Gamma
(\varepsilon ),  \label{sigma}
\end{equation}%
By setting $E=0$ in the whole system (i.e. setting down the Fermi level as
the energy reference) the analysis is further simplified. Equation \ref%
{sigma} has two solutions with different signs. The solution with physical
meaning provides a retarded response and results: 
\begin{equation}
\Sigma (\varepsilon )=\dfrac{\varepsilon +\mathrm{i}\eta }{2}-\mathrm{sgn}%
(\varepsilon )\times \left( \sqrt{\dfrac{r+x}{2}}+\mathrm{i\times sgn}%
(y)\times \sqrt{\dfrac{r-x}{2}}\right) ,  \label{e4}
\end{equation}%
with $x=\dfrac{\varepsilon ^{2}-\eta ^{2}}{2}-V^{2}$, $y=\dfrac{\varepsilon
\eta }{2}$ and $r=\sqrt{x^{2}+y^{2}}$.

Then, the restriction to the first four orbitals of the total Hamiltonian
can be written in a simple way: 
\begin{equation}
\widetilde{\mathbb{H}}=\left[ 
\begin{array}{cccc}
\Sigma ^{z^{2}}(\varepsilon ) & -V_{0} & -V_{0} & 0 \\ 
-V_{0} & -\mathrm{i}\eta & -V_{AB} & +\lambda V_{0} \\ 
-V_{0} & -V_{AB} & -\mathrm{i}\eta & -\lambda V_{0} \\ 
0 & +\lambda V_{0} & -\lambda V_{0} & \Sigma ^{xz}(\varepsilon )%
\end{array}%
\right] .  \label{e5}
\end{equation}

Now, a basis change can be made to a molecular \textit{bonding} and \textit{%
antibonding} representation. Equation \ref{e6} shows the Hamiltonian in the
new basis. Notice that, the bonding state (second diagonal element) does not
interact with $\Sigma ^{xz}(\varepsilon )$ (fourth diagonal element) and the
antibonding state (third diagonal element) does not interact with $\Sigma
^{z^{2}}(\varepsilon )$ (first diagonal element):

\begin{equation}
\widetilde{\mathbb{H}}^{\prime }=\widetilde{\mathbb{H}}_{\mathrm{+}}\otimes 
\widetilde{\mathbb{H}}_{\mathrm{-}}=\left[ 
\begin{array}{cccc}
\Sigma ^{z^{2}}(\varepsilon ) & -\sqrt{2}V_{0} & 0 & 0 \\ 
-\sqrt{2}V_{0} & -V_{AB}-\mathrm{i}\eta & 0 & 0 \\ 
0 & 0 & V_{AB}-\mathrm{i}\eta & \sqrt{2}\lambda V_{0} \\ 
0 & 0 & \sqrt{2}\lambda V_{0} & \Sigma ^{xz}(\varepsilon )%
\end{array}%
\right] .  \label{e6}
\end{equation}%
Therefore, the system is naturally detached in two portions in which the
Green's function matrices can be solved independently. For the bonding
subspace, i.e. the bonding molecular orbital interacting with $\Sigma
^{z^{2}}(\varepsilon )$, the Green's function takes the form:

\begin{equation}
\mathbb{G}_{+}=\dfrac{1}{(\varepsilon +V_{AB}+\mathrm{i}\eta )(\varepsilon
-\Sigma ^{z^{2}}(\varepsilon ))-2V_{0}^{2}}\left[ 
\begin{array}{cc}
\varepsilon +V_{AB}+\mathrm{i}\eta & -\sqrt{2}V_{0} \\ 
-\sqrt{2}V_{0} & \varepsilon -\Sigma ^{z^{2}}(\varepsilon )%
\end{array}%
\right] ,  \label{e7}
\end{equation}%
while, for the antibonding molecular orbital interacting with $\Sigma
^{xz}(\varepsilon )$, there is a subspace where

\begin{equation}
\mathbb{G}_{-}=\dfrac{1}{(\varepsilon -V_{AB}+\mathrm{i}\eta )(\varepsilon
-\Sigma ^{xz}(\varepsilon ))-2(\lambda V_{0})^{2}}\left[ 
\begin{array}{cc}
\varepsilon -\Sigma ^{xz}(\varepsilon ) & \sqrt{2}\lambda V_{0} \\ 
\sqrt{2}\lambda V_{0} & \varepsilon -V_{AB}+\mathrm{i}\eta%
\end{array}%
\right] .  \label{e8}
\end{equation}

For the rest of the article $\lambda$ will be set $\lambda \sim 1$ and $%
\Sigma ^{xz}=\Sigma ^{z^{2}}$. The eigenenergies and resonances of the
system are obtained by finding the poles of Eqs. \ref{e6} and \ref{e7}. This
is achieved solving the equations:

\begin{eqnarray}  
\label{Eq-CharacBonding}
\varepsilon +V_{AB}-2\alpha \Sigma (\varepsilon )=0, \\
\label{Eq-CharacAntibonding}
\varepsilon -V_{AB}-2\alpha \Sigma (\varepsilon )=0.
\end{eqnarray}
Equation \ref{Eq-CharacBonding} accounts for the poles corresponding to the
bonding state interacting with the $d_{z^{2}}$ band and Eq. \ref%
{Eq-CharacAntibonding} for the poles of the antibonding state interacting
with the $d_{xz}$ band.

\section{Molecular dissociation}

\begin{figure}[tbh!]
\centering{\ \includegraphics[width=8cm]{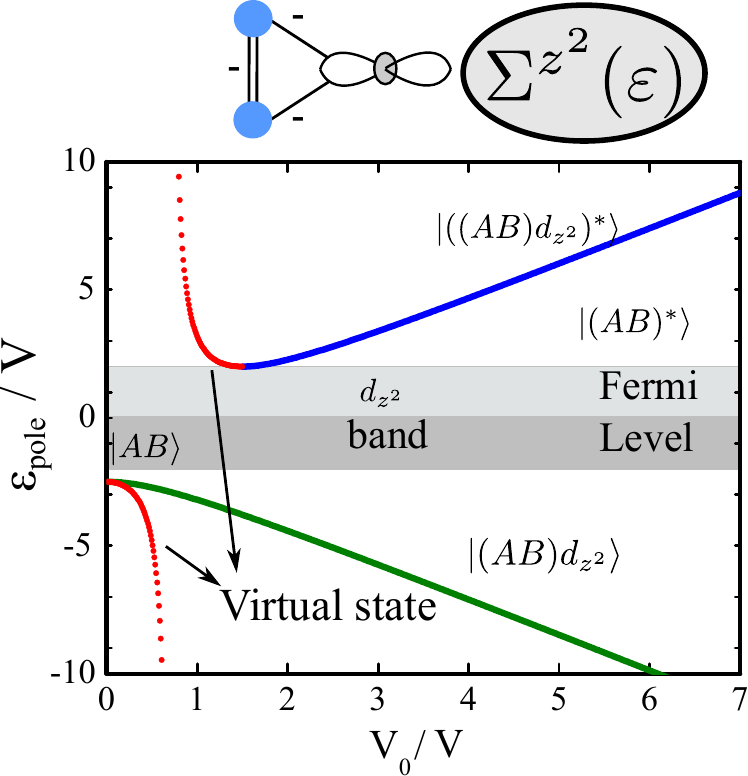} }
\caption{Poles of the Green's function for the parallel configuration when
the molecule interacts with the $d_{z^2}$ orbital. }
\label{polo+}
\end{figure}

A first hint for molecular dissociation arises from analysing the molecular
bonding orbital that intercats with the $d$ band through the $d_{z^{2}}$
orbital, Fig. \ref{polo+}. In this case, Eq. \ref{Eq-CharacBonding} provides
two poles which are bellow the $d$ band at the molecular bonding energy $%
\varepsilon =-V_{AB}$. One is a physical localized pole (green line in Fig. %
\ref{polo+}) which corresponds to the bonding state $\left\vert
AB\right\rangle $. As the interaction increases, $\left\vert AB\right\rangle 
$ evolves to a bonding combination between the bonding state of the molecule
and the metal, i.e. $\left\vert (AB)d_{z^{2}}\right\rangle $, becoming more
localized and its energy lying well below the Fermi level. The other pole
corresponds to a non-physical virtual state which, as the interaction
increases, escapes to negative energies and reappears at positive values
(red dots in Fig. \ref{polo+}). As the non-physical pole gets closer to the $%
d$ band, it finally meets the band-edge and suffers a transition into a
physical localized state. This is an antibondig combination between the
molecular bonding state and the metal $\left\vert ((AB)d_{z^{2}})^{\ast
}\right\rangle $ (blue line). In this scenario, bound weakening occurs
because occupying the $\left\vert (AB)d_{z^{2}}\right\rangle $ state implies
diminish the occupation of the bonding $\left\vert AB\right\rangle $ from
100\% into a final 50\%. Indeed, the molecular bonding state now has 50\%
participation in the unoccupied $\left\vert ((AB)d_{z^{2}})^{\ast
}\right\rangle $ localized orbital that emerged from the upper top of the $d$
band.

\begin{figure}[tbh]
\centering{\ \includegraphics[width=8cm]{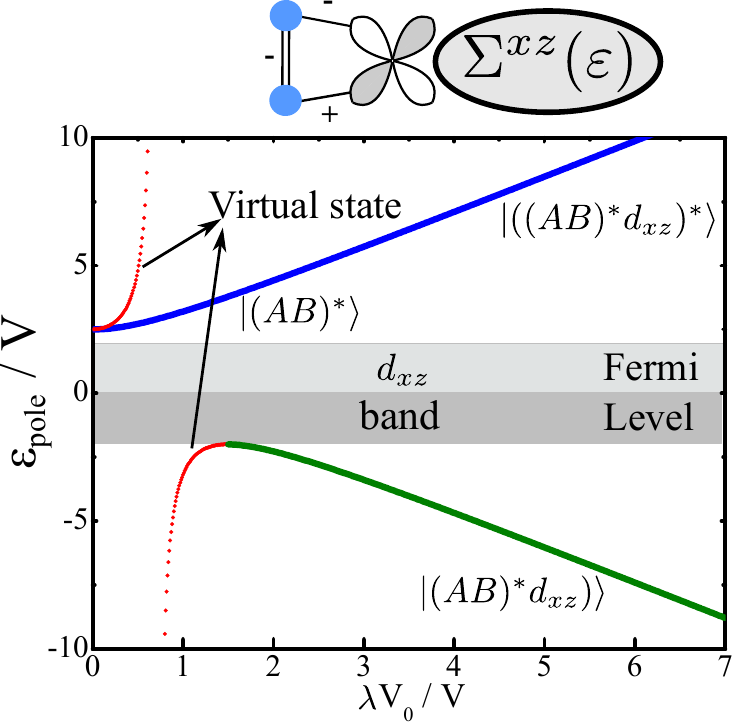} }
\caption{Poles of the Green's function for the parallel configuration when
the molecule interacts with the $d_{xz}$ orbital. The molecule dissociation
as a QDPT can be observed when the interaction is with the $d_{xz}$ band.}
\label{polo-}
\end{figure}

The previous discussion has a precise equivalence in the analysis of the
states that evolve from the molecular antibonding state. However, the same
formulation has now completely different meaning. The molecular antibonding
state interacts with the $d$ band through $d_{xz}$. The poles resulting from
Eq. \ref{Eq-CharacAntibonding}, are shown in Fig. \ref{polo-}. At the
antibonding energy $\varepsilon =V_{AB}$, two poles appear. A physical
localized state, related to the molecular antibonding state $\left\vert
(AB)^{\ast }\right\rangle $ (blue line in Fig. \ref{polo-}), whose energy
increases as $V_{0}$ increases and becomes an antibonding combination
between the molecular antibonding state and the metal site $%
\left\vert((AB)^{\ast }d_{xz})^{\ast }\right\rangle $. The other pole at $%
\varepsilon =V_{AB}$ is a virtual state \cite{Dente,bustos2010buffering}
(red dots in Fig. \ref{polo-}) which diverges as $V_{0}$ increases and
appears again from $-\infty$ until its energy touches the $d$ band. At this
critical value, the virtual state suffers a transition and becomes a
localized state (green line in Fig. \ref{polo-}) which is a bonding
combination between the molecular antibonding state and the metal band $%
\left\vert (AB)^{\ast }d_{xz}\right\rangle $. Therefore, molecular
dissociation can be interpreted as occurring at the precise value when the
virtual pole touches the $d$ band and becomes the localized, and occupied,
state$\left\vert(AB)^{\ast }d_{xz}\right\rangle$. Thus, molecular
dissociation occurs at a non-analytical point of the physical observables,
e.g. total energies. At this point the molecular electrons have a transition
from an increasingly occupied bonding state that participates of the
delocalized band into a localized combination between the $d$ states and
antibonding molecular orbital. This is a form of Quantum Dynamical Phase
Transition which, to the best of our knowledge, has not been identified
before in the context of molecular dissociation.

\begin{figure}[tbh]
\centering{\ \includegraphics[width=10cm]{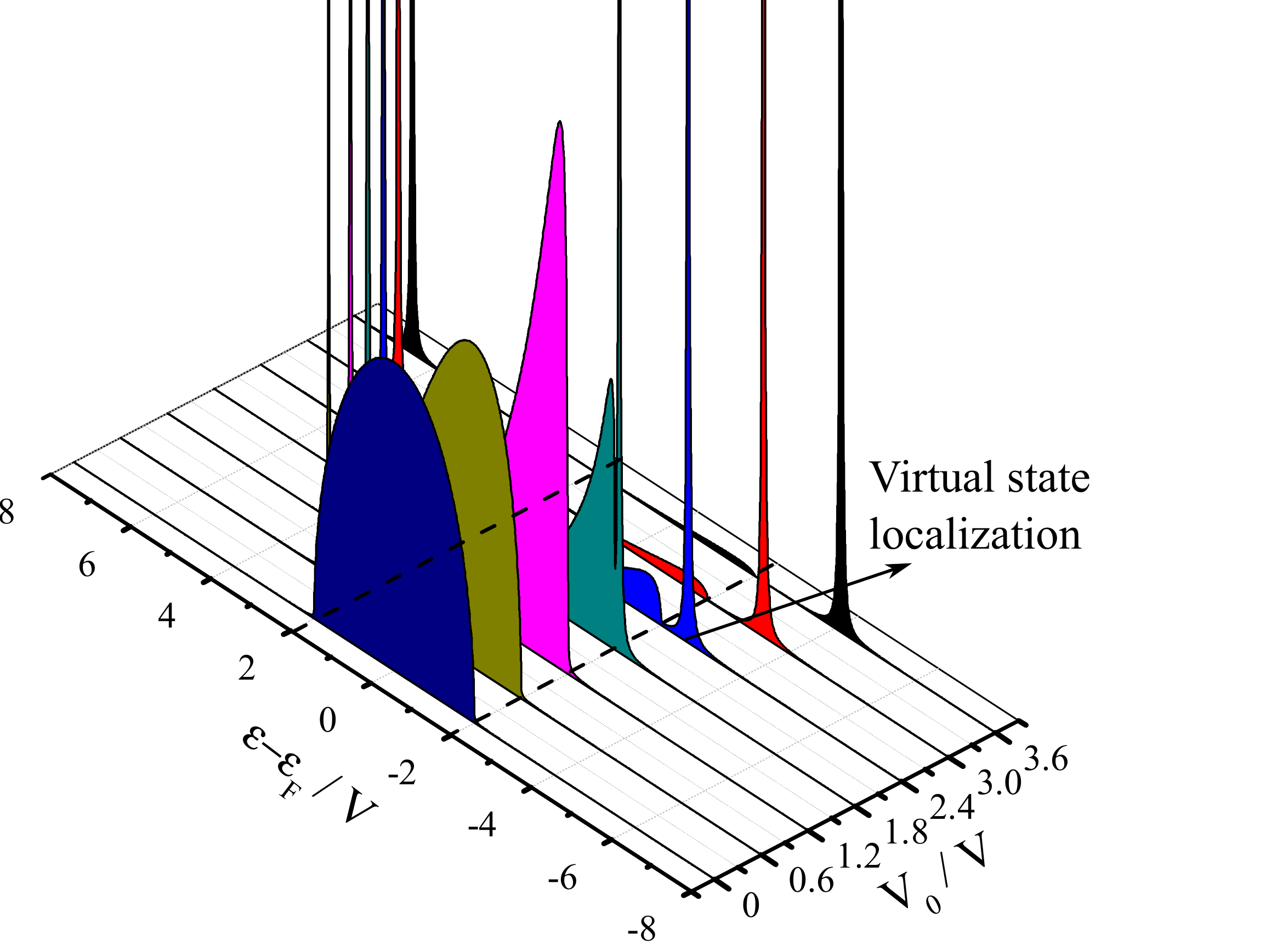} }
\caption{LDoS of the $d$ band. As $V_{0}$ increases a state is expelled from
the band and, after the transition point, forms the localized state $%
\left\vert (AB)^{\ast }d_{xz}\right\rangle $, $\protect\eta =0.01$ eV.}
\label{dens}
\end{figure}

From the results it becomes evident that the most interesting situation is
when the antibonding molecular orbital interacts with the $d_{xz}$. From Eq. %
\ref{e8} we get the diagonal Green's function at the $d_{xz}$ metallic
orbital:

\begin{equation}
G_{d_{xz}}(\varepsilon )=\dfrac{1}{\varepsilon +\mathrm{i}\eta -\Sigma
(\varepsilon )-\dfrac{2(\lambda V_{0})^{2}}{\varepsilon +\mathrm{i}\eta
+V_{AB}}}.  \label{e11}
\end{equation}%
The LDoS for the $d$ band can be obtained from Eq. \ref{e11},

\begin{equation}
N_{d_{xz}}(\varepsilon )=-\dfrac{1}{\pi }\lim_{\eta \rightarrow 0^{+}}%
\mathrm{Im}\left[ G_{d_{xz}}(\varepsilon )\right] ,  \label{e12}
\end{equation}%
which becomes of great help to reinforce and extend the previous discussion.
This LDoS is shown in Fig. \ref{dens} for $\lambda V_{0}/V$ between 0 and
3.6 for $\lambda=1$. When $V_{0}\sim 0$ the shape of the LDoS corresponds to
a non interacting $d_{xz}$ band. As $\lambda V_{0}$ increases the $d_{xz}$
band starts to mix with the antibonding state of the dimer. The energy of
this antibonding combination $\left\vert ((AB)^{\ast }d_{xz})^{\ast
}\right\rangle$, progresses toward increasingly positive values as the
interaction grows. Meanwhile, the virtual state approaches the $d_{xz}$ band
from negative energies while it produces an \textquotedblleft
attraction\textquotedblright\ that increases the LDoS near the band edge. As
the virtual state meets the band a localized state \textit{emerges} from the
band edge and gains weight. A similar issue was recently discussed in the
context of engineered plasmonic excitations in metallic nanoparticle arrays 
\cite{bustos2010buffering}. There, it was shown analytically that the
distorted band is the product among the original semi-elliptic band and a
Lorentzian centered in the virtual state. This concentrates a density of
states near band edge until it becomes a divergence and a localized state is
expelled at a critical interaction strength, shown as a dot in Fig. \ref%
{polo-}.

\begin{figure}[tbh]
\centering{\ \includegraphics[width=10cm]{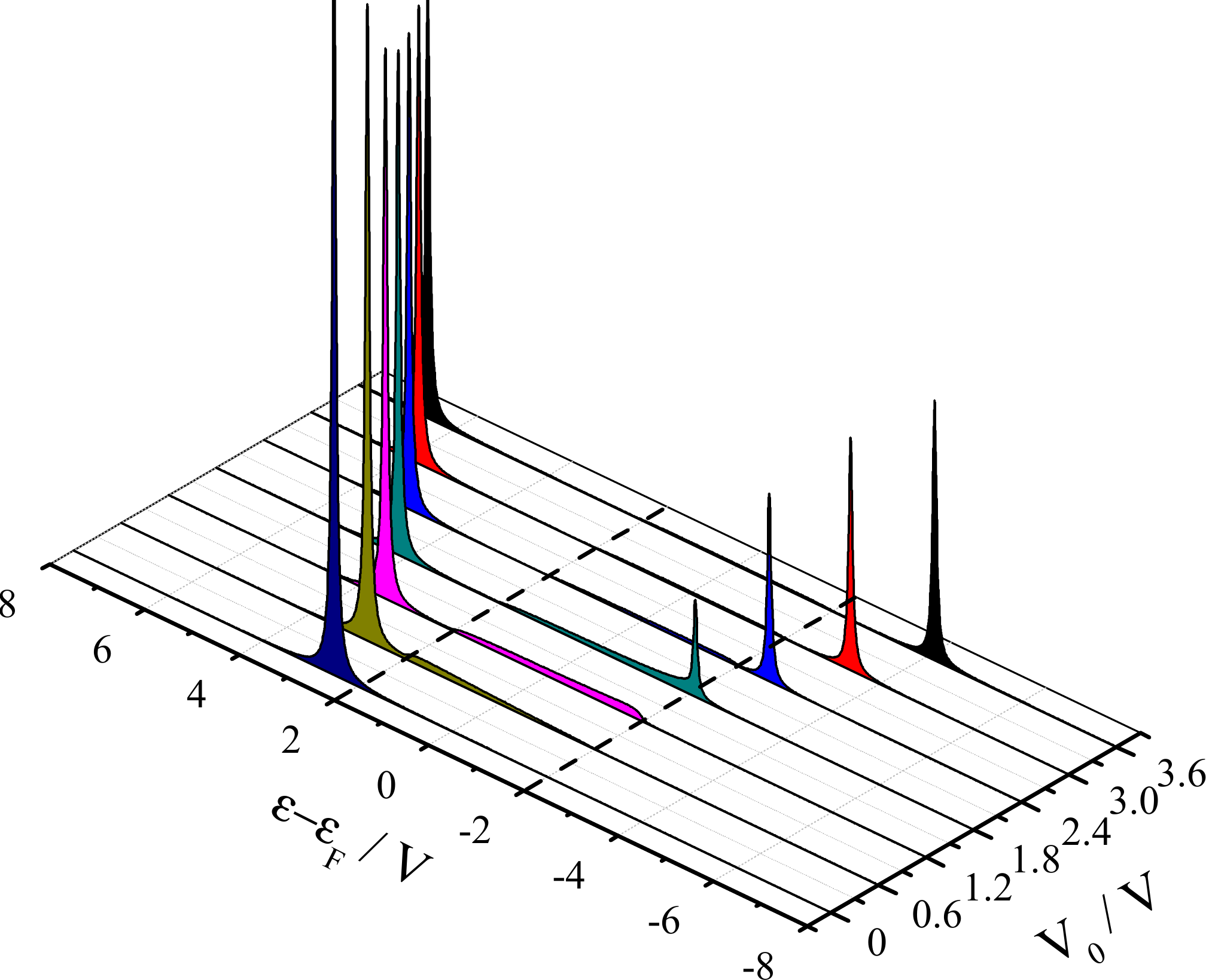} }
\caption{LDoS of the molecular antibonding state $\left\vert (AB)^{\ast
}\right\rangle$, interacting with the metallic orbital $d_{xz}$, as $V_{0}$
increases, $\protect\eta =0.05$ eV.}
\label{dens2}
\end{figure}

The previous conclusion is reinforced by the analysis of LDoS at the
antibonding orbital. Figure \ref{dens2} shows how the unoccupied antibonding
state $\left\vert (AB)^{\ast }\right\rangle $ looses its weight towards a
participation on combination with the $d_{xz}$ band which finally emerges as
an \textit{occupied localized state}. This is a crucial contribution to
molecular destabilization. As in the first part of this work \cite{grado1}
the new transition can be seen as a successful implementation of a
non-Hermitian Hamiltonian \cite{rotter2015review} in a well defined model.

Notice that Figs. \ref{dens} \ and \ref{dens2} also serve to discuss the
interaction between the bonding molecular state $\left\vert AB\right\rangle $
and the $d_{z^{2}}$ band by exchanging the sign of the energy. Thus, in this
case, the $\left\vert ((AB)d_{z^{2}})^{\ast }\right\rangle $ emerges as an
unoccupied localized state above the $d_{z^{2}}$ band, while $\left\vert
AB\right\rangle $ state loses occupation as the $\left\vert
(AB)d_{z^{2}}\right\rangle $ state forms with increasing interaction.

\section{Conclusions}

As a H$_{2}$ molecule approaches a catalyst with its axis parallel to the
surface, the interaction creates two independent collective orbitals which
are superpositions with different surface $d$ orbitals that are part of
their corresponding metallic bands. The molecular bonding state becomes
mixed with the $d_{z^{2}}$ band while the molecular antibonding state
interacts with the $d_{xz}$ band. This gives rise to two processes described
by the same algebra, but with different physical meanings as their energies
are the reverse of each other.

On one side, the mixing of the molecular bonding state produces a decrease
of its occupation. While this occurs, the LDoS of the $d_{z^{2}}$ band is
distorted at its upper edge much as if it were \textquotedblleft
attracted\textquotedblright\ upwards. Finally, at a critical interaction
strength the divergent peak is expelled as a localized state emerging from
the upper (i.e. unoccupied) part of the $d_{z^{2}}$ band. This new \textit{%
unoccupied} state is an antibonding combination among the surface $d_{z^{2}}$
orbital and the \textit{bonding} state of the dimer.

On the other side, a fraction of the molecular antibonding state gets
increasingly mixed with the $d_{xz}$ metallic band. This produces a decrease
of the dimer participation on its unoccupied antibonding combination.
Simultaneously, the $d_{xz}$ \ LDoS is \textquotedblleft
attracted\textquotedblright\ towards its lower edge until it finally emerges
as an \textit{occupied localized state} build as a bonding combination among
the molecular \textit{antibonding} state and the $d_{xz}$ band.

\begin{figure}[tbp]
\begin{equation*}
\xymatrix{ &&&\left\vert((AB)d_{z^{2}})^*\right\rangle&\\ \left\vert
AB\right\rangle\ar @{<->} [r] &d_{z^2}\ar[urr]\ar[drr]&&&\txt{$\left\vert
AB\right\rangle$ 50 \% unoccupied}\\ &&&\left\vert(AB)d_{z^{2}}\right\rangle
& \\ &&&\left\vert((AB)^*d_{xz})^*\right\rangle& \\
\left\vert(AB)^*\right\rangle \ar @{<->} [r]&
d_{xz}\ar[urr]\ar[drr]&&&\txt{$\left\vert(AB)^*\right\rangle$ 50 \%
occupied}\\ &&&\left\vert(AB)^*d_{xz}\right\rangle& }
\end{equation*}%
\caption{The interaction of the bonding molecular orbital with the $%
d_{z^{2}} $ band shields an antibondig combination that depopulates this
molecular orbital, while the occupied fraction losses weight towards the $%
d_{z^{2}}$ band. Simultaneously, the interaction of the antibonding
molecular orbital with $d_{xz}$ band enforces this molecular state to split
among an antibonding combination and an emergent bonding one that is
interpreted as the molecular breakdown. }
\label{figura7}
\end{figure}
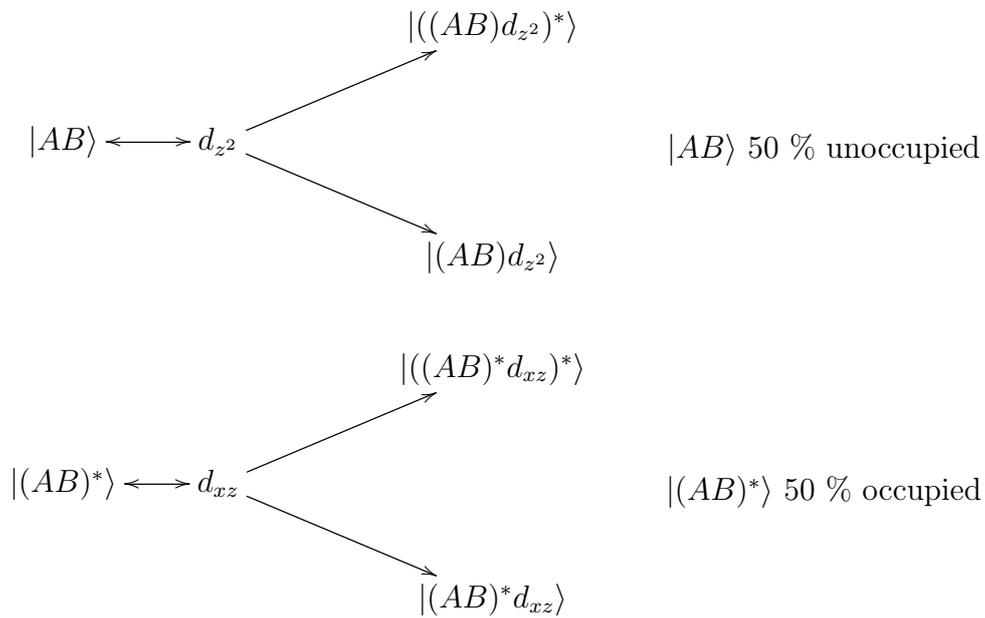

These simultaneous mixing processes, i.e. the depopulation of the molecular
bonding state and the occupation of the molecular antibonding state, both
schematized in Fig. \ref{figura7}, are responsible for the dimer
destabilization that leads to its breakdown.

While the essence of the molecule dissociation mechanisms are already hinted
by the resolution of toy models for the catalyst such as small metallic
clusters or even a single metal atom, the criticality of the dissociation
transition would not be readily captured. Indeed, as in the first part of
this work \cite{grado1}, the quasi-continuum nature of a metallic substrate
is crucial to describe dissociation as an analytical discontinuity. In this
case, we interpreted dissociation as the emergence of the localized state
from the band edges as the interaction strength increases. This is an actual
quantum dynamical phase transition. Remarkably, the elusive virtual states
(i.e. states that are non-physical poles of $\mathbb{G}(\varepsilon )$ \cite%
{landau1965course, moiseyev}) acquire a physical meaning as
\textquotedblleft attractors\textquotedblright\ of a distortion of the
continuum band creating a LDoS divergence that finally expels a localized
state. This is, a non-analytical transition. 

\section*{Acknowledgements}

We acknowledge the financial support from CONICET (PIP 112-201001-00411),
SeCyT-UNC, ANPCyT (PICT-2012-2324) and DFG (research network FOR1376). We
thank P. Serra and W. Schmickler for discussions and references.

\newpage

\bibliographystyle{jpc}

\end{document}